\documentclass[aps,prb,twocolumn,showpacs,superscriptaddress,groupedaddress]{revtex4-1}

\usepackage{graphicx}
\usepackage{amssymb,bbm}
\usepackage{amsmath}
\newcommand{\beq}{\begin{equation}}
\newcommand{\eeq}{\end{equation}}

\newcommand{\<}{\left<}
\renewcommand{\>}{\right>}
\newcommand{\aJ}{a_{\!J}}

\newcommand{\uu}[1]{{\boldsymbol #1}}

\def\mb{\uu{m}}
\def\Pb{\uu{m}_p}
\def\etab{\uu{\eta}}

\def\heff{\uu{h}_{\text{eff}}}

\begin{document}

\title{Averaged equation for energy diffusion on a graph reveals bifurcation diagram \\and thermally assisted reversal times in spin-torque driven nanomagnets}

\author{Katherine A. Newhall}
\affiliation{Courant Institute of Mathematical Science, New York University, New York, New York 10012}

\author{Eric Vanden-Eijnden}
\affiliation{Courant Institute of Mathematical Science, New York University, New York, New York 10012}

\date{\today}

\begin{abstract}
  Driving nanomagnets by spin-polarized currents offers exciting prospects in magnetoelectronics, but the response of the magnets to such currents remains poorly understood.  We show that an averaged equation describing the diffusion of energy on a graph captures the low-damping dynamics of these systems. From this equation we obtain the bifurcation diagram of the magnets, including the critical currents to induce stable precessional states and magnetization switching, as well as the mean times of thermally assisted magnetization reversal in situations where the standard reaction rate theory of Kramers is no longer valid. These results match experimental observations and give a theoretical basis for a N\'eel-Brown-type formula with an effective energy barrier for the reversal times.  \end{abstract}



\maketitle

\section{Introduction}

Manipulating thin-film magnetic elements with spin-polarized currents besides external magnetic fields~\cite{KSKESBR2003} has generated a lot of recent interest in applications to magnetoelectronic devices that offer low power memory storage without the use of moving parts~\cite{IEEE2012}. Understanding the response of the magnet to such currents is nontrivial, however, because they apply a nonconservative force, called spin-transfer torque (STT), on the system. Like other nongradient systems with no Lyapunov function, the phase portrait of nanomagnets in the presence of STT can be quite complex, and include limit cycles or chaotic trajectories besides fixed points. When the applied fields and/or currents are nonstationary, and in the presence of thermal noise, the situation is even worse. In particular, Kramers' reaction rate theory~\cite{Kramers1940,Review_Kramers} is no longer applicable and the N\'eel-Brown formula~\cite{Brown1963} for the mean magnetization reversal time is not valid since there is no well-defined energy associated with STT.

Nanomagnets typically operate in a regime where the nonconservative parts of the dynamics, including the effects of damping, STT, and thermal noise, act on time-scales that are much longer than that of the energy-conserving Hamiltonian part. Trajectories remain close to periodic Hamiltonian orbits for a long time, and slowly drift from one orbit to another due to damping, STT, and thermal noise.  This separation of time scales can be exploited, using averaging techniques developed by Freidlin and Wentzell~\cite{FWbook,FH2011} (see also~[\onlinecite{Multiscale}] and [\onlinecite{NanoMagBook}]), to reduce the dynamics to that of an energy diffusing on a graph. We show here that this reduced description permits to explain the features of nanomagnets subject to STT that are observed experimentally.  Specifically, we obtain the full bifurcation diagram of the system at zero temperature and determine the critical spin-polarized currents needed to induce stable precessional states~\cite{KEGSKRB2004,Bazaliy2007} and magnetization switching~\cite{KEGSKRB2004,KMSRK2003}.  At finite temperature, we calculate the mean times of thermally assisted magnetization reversals~\cite{KEGSKRB2004,MASBBR2002}, and give expressions for the effective energy barriers conjectured to exist~\cite{KEGSKRB2004,MASBBR2002,AV2005,LZ2003,LZ2004}. These results are complementary to those obtained in~[\onlinecite{CSKV2011}], using the geometrical Minimum Action Method (gMAM)~\cite{ERV2004,gmam}, for situations with small thermal noise and stronger damping.

The remainder of this paper is organized as follows.  In Sec.~\ref{sec:modeling} we present the governing equation and briefly explain the origin of the terms.  In Sec.~\ref{sec:average} we discuss in detail the separation of time scales underlying the averaging procedure and obtain the reduced description of energy diffusing on a graph.  In Sec.~\ref{sec:asym_min} we obtain asymptotic approximations for the averaged coefficients near the energy minimum, and in Sec.~\ref{sec:asym_sad} we do the same near the saddle point in energy.  We present the full bifurcation diagram of the system in Sec.~\ref{sec:resultsA} and give expressions for the effective energy barriers within the mean times of thermally assisted magnetization reversals in Sec.~\ref{sec:resultsB}.  Some conclusions are presented in Sec.~\ref{sec:conclusions} and technical details are deferred to Appendices.

\section{Modeling Equation}
\label{sec:modeling}

We will focus on magnetic systems in which the magnetization has constant strength $M_s$ in the direction of a unit vector $\mb(t) = (m_x (t), m_y (t), m_z (t))$ whose evolution is governed by 
\begin{equation}
  \label{eq:system} 
  \dot\mb=-\mb\times \heff +  {\bf m } \times \left(\mb\times 
    \left(-\alpha \heff + a_{\!J} \Pb\right)\right).
\end{equation} 
This is the standard stochastic Landau-Lifshitz-Gilbert (LLG) equation written in non-dimensional form with an additional STT term, $\mb\times (\mb\times a_{\!J}{\Pb})$, modeling the transfer of angular momentum to the magnetization from the electron spin in a polarized current directed along unit vector $\Pb$.  The non-dimensional current strength~\cite{RS2008,Nature2012}, 
\begin{equation}\label{eq:aJ}
\aJ=I \frac{\gamma^*\mu_0 \eta \mu_B}{ e\nu},
\end{equation}
contains the dependence on the electrical current, $I$, and the structural and material properties of the fixed and free magnetic layers through $\eta$.  The division by the volume, $\nu$, of the free layer produces a force per volume, matching the energy per volume contribution already in~\eqref{eq:system}.  For simplicity, here we take $\Pb=(1,0,0)$ and a constant strength $a_{\!J}$, but these could straightforwardly be generalized to any direction and a time-varying strength.  

The other terms in \eqref{eq:system} are standard.  The parameter $\alpha$ is the non-dimensional Gilbert damping parameter. The effective magnetic field, divided by $\mu_0M_s$, is the non-dimensional term
\begin{equation}
\heff=-\nabla_m E + \sqrt{\frac{2\alpha\epsilon }{ 1+\alpha^2}}\etab(t)\label{eq:1}
\end{equation}
which in turn is the sum of the negative gradient of the non-dimensional energy per volume, $E(\mb)$, and a term accounting for thermal effects with $\etab(t)$ being three-dimensional white-noise.   The noise amplitude, $\sqrt{2\alpha\epsilon/(1+\alpha^2)}$ is consistent with the equilibrium distribution being the Gibbs distribution (as shown in Appendix A of~[\onlinecite{KGV2005}]) and 
\begin{equation}
\epsilon = k_B T /\mu_0M_s^2\nu\label{eq:2}
\end{equation}
is the non-dimensional temperature.  The energy per volume, divided by $\mu_0M_s^2$, is chosen here to have the form  
\begin{equation}
E(\mb) = \beta_y m_y^2+\beta_z m_z^2 - h_xm_x \label{eq:3}
\end{equation}
with $\beta_y = H_k/2M_s < \beta_z=1/2$, corresponding to biaxial anisotropy, $H_k$, along the $y$ direction.  Additionally, a magnet with non-dimensional energy
$$ E(\mb) = \beta_x(1-m_x^2) - h_xm_x $$
corresponding to uniaxial anisotropy is considered in Appendix~\ref{sec:uni}. A planar applied field of amplitude $\mu_0 M_s h_x$ is applied along the $x$-direction.  Last, we point out that in~\eqref{eq:system} time has been nondimensionalized by $\gamma\mu_0 M_s/(1+\alpha^2)$ where $\gamma$ is the gyromagnetic ratio.

While we work in terms of nondimensionalized equations throughout most of the paper, in Sec.~\ref{sec:results} when we present our results, we use dimensional variables to facilitate comparison with experimental results.  To this end, we use a set of parameters representative of a typical ferromagnet used in experiments:  the saturation magnetization $\mu_0 M_s=1.2$T and damping parameter $\alpha=1.5\times 10^{-3}$ are taken from [\onlinecite{ChenEtAl2007}], the gyromagnetic ratio is $\gamma = 2.21\times 10^5$m/As, the parameter $\beta_y=0.015$ is chosen to match the hysteresis curve in  [\onlinecite{KEGSKRB2004}], and the non-dimensional temperature $\epsilon=0.004$ corresponds to a temperature $T=300K$ and a volume of $12\times 12 \times 2$nm${}^3$, which is reasonably described by a single magnetic vector.  

\section{Averaged Energy Equation}
\label{sec:average}

\begin{figure}
   \centering
   \includegraphics{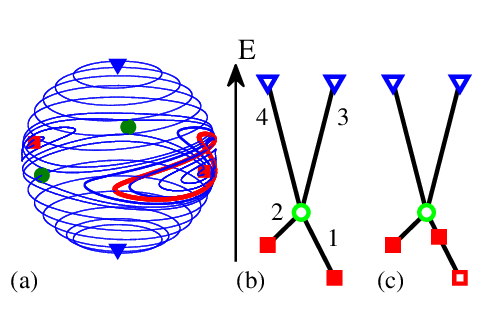}
   \caption{(color online) (a): Hamiltonian orbits (blue lines) of the unit magnetization vector solution of $\dot{\mb}=-\mb\times \nabla_m E $ along with fixed points of this equation that are also critical points of the energy: minima (red squares), saddle points with energy $E_b$ (green dots), and maxima (blue triangles).  (b) and (c): Graphs (not drawn to scale) in which each energy point along the edges corresponds to an orbit of $\mb$ shown in (a) with this energy.  The numbers indicate the label for each energy branch.  Branch 1: $m_x> -h_x/2\beta_y$ and $E<E_b$, branch 2: $m_x< -h_x/2\beta_y$ and $E<E_b$, branch 3: $m_z>0$ and $E>E_b$ and branch 4: $m_z<0$ and $E>E_b$.  The ends of each branch correspond to the fixed points in (a) and the markers indicate the location and stability of the fixed points of $\dot{E}=-\alpha A_j(E)+a_{\!J}B_j(E)$: filled markers are stable and open markers are unstable.  The graph in (b) corresponds to a situation where the energy minima are stable fixed points. The graph in (c) is a representative case when a solution of~\eqref{eq:E0} exists, leading to a new stable fixed point at $E=E_0$; this new fixed point corresponds to a stable limit cycle like the one shown as a thick red line in (a).}
   \label{fig:orbits}
\end{figure}

The presence of the STT term in~\eqref{eq:system}, which is nongradient and nonconservative, complicates the analysis of this equation even in the absence of thermal noise ($\epsilon=0$). In particular, the magnetic energy $E(\mb)$ is not a Lyapunov function for the system. Understanding the effect of the STT term is a question that has received much attention in both theoretical and experimental literatures~\cite{RS2008,KSKESBR2003,KEGSKRB2004,WB2002,Bazaliy2007,KMSRK2003,MASBBR2002,AV2005,LZ2003,LZ2004,CSKV2011}.  Here, we address this question by taking advantage of the separation of time scales that arises when both the damping and the strength of the polarized current are weak, $\alpha\ll1$ and $ a_J\ll 1$. In this regime, $\mb$ moves rapidly along the energy conserving Hamiltonian orbits in Fig.~\ref{fig:orbits}(a) and drifts slowly in the direction perpendicular to these orbits. This slow motion can be captured by tracking the evolution of the energy~$E(\mb)$ along with an index to distinguish between disconnected orbits with the same energy. This information is encoded in the graph shown in Fig.~\ref{fig:orbits}(b), whose topology is directly related to the energy function, $E=\beta_ym_y^2+\beta_zm_z^2-h_xm_x$, and changes based on its form and values of parameters.  For example, when $|h_x|< 2 \beta_y$ the graph has four branches, as shown in Fig.~\ref{fig:orbits}(b), which meet at the saddle point of the energy that corresponds to the homoclinic orbits connecting the two green points on the surface of the sphere in Fig.~\ref{fig:orbits}(a).  We will use the indexes 1 and 2 (3 and 4) for the lower (higher) energy branches in Fig.~\ref{fig:orbits}(b), which correspond to orbits on the front-right and back-left (top and bottom) of the sphere in Fig.~\ref{fig:orbits}(a), respectively.

To deduce the effective dynamics on the graph when $\alpha$ and $\aJ$ are small, we follow Freidlin and Wentzell~\cite{FWbook} to remove the direct dependence on $\mb(t)$ from $\dot{E} = \nabla_m E \cdot\dot \mb$.
First, we convert \eqref{eq:system} to an Ito SDE, then determine $\dot{E}$ using the stochastic chain rule (details in Appendix~\ref{app:Ito}).  To remove the explicit dependence on the magnetization vector $\mb$  from the equation for $\dot{E}$, we average the coefficients appearing in the backwards Kolmogorov equation for the SDE of $\dot{E}$  over one period, $T_j(E)$, at constant energy,
\begin{equation}\label{avg_fnc}
\<f(\mb)\>_j=\frac{1}{T_j(E)}\int_0^{T_j(E)} f\big(\mb(t)\big) dt.
\end{equation}
The subscript $j=1,2,3,4$ indicates that the average corresponds to one connected orbit of $\mb$ with constant energy~$E$ on branch~$j$ of the energy graph (see Fig.~\ref{fig:orbits}(b)).  The resulting averaged coefficient backwards Kolmogorov equation corresponds to the averaged coefficient SDE
\begin{equation}
  \label{eq:avg_E}
  \dot{E} = -\alpha A_j(E)+ a_{\!J} B_j(E) + 2\alpha\epsilon C_j(E) +
  \sqrt{2\alpha\epsilon A_j(E)}\xi(t), 
\end{equation}
written in Ito's form, where $\xi(t)$ is a 1D white-noise and 
\begin{equation}
  \label{eq:coeff}
  \begin{aligned} 
    A_j(E)&= 4\big(\beta_y^2\<m_y^2\>_j+\beta_z^2\<m_z^2\>_j - E^2\big)\\
    &  \quad- 4Eh_x\<m_x\>_j  +  h_x^2\big(1-\<m_x^2\>_j\big)\\
    B_j(E) &=  2E\<m_x\>_j + h_x\big(1+\<m_x^2\>_j\big) \\
    C_j(E)&= \beta_y+\beta_z - 3E- 2h_x\<m_x\>_j \;.
  \end{aligned}
\end{equation}   
In Appendix~\ref{sec:uni} we derive the coefficients in \eqref{eq:coeff} for the alternate case when the energy corresponds to uniaxial anisotropy.

As we will show next in Secs.~\ref{sec:asym_min} and \ref{sec:asym_sad}, the averages in~\eqref{eq:coeff} can be evaluated asymptotically near the critical points.  This information turns out to be sufficient to calculate the bifurcation diagram and the mean times of magnetization reversal that we obtain in Sec.~\ref{sec:results}. Away from the critical points, the averages~\eqref{eq:coeff} must be evaluated numerically, which we do by using a symplectic implicit mid-point integrator to evolve $\mb$ via $\dot \mb = \mb \times \nabla_m E$ along an orbit with prescribed energy to compute the time averages.  Note also that (\ref{eq:avg_E}) requires a matching condition where the branches on the energy graph meet~\cite{FWbook}; these conditions are discussed in Appendix~\ref{sec:Matching}.

\subsection{Approximation Near the Minima\label{sec:asym_min}}

In order to determine the scaling of the averaged coefficients $A_1(E)$, $B_1(E)$ and $C_1(E)$, near the energy minimum on branch 1, $\mb_0=(1,0,0)$, we create a series expansion about this point for the solution to the Hamiltonian system,
\beq\label{HamSys}
\dot{\mb} = -\mb \times \nabla E = \left( \begin{array}{c}
2(\beta_z-\beta_y)m_ym_z \\ -m_z(2\beta_z m_x+h_x) \\ m_y(2\beta_y m_x + h_x) 
\end{array}\right),
\eeq
then compute the averages exactly as a function of the energy.  This expansion must also satisfy the constraint that $|\mb|^2=1$.  Utilizing standard perturbation methods, we obtain the expansion
$$\begin{aligned}
m_x & \sim 1 - \delta^2 \frac{1}{2}\Big[ \cos^2\omega t + \frac{2\beta_y + h_x}{2\beta_z + h_x}\sin^2 \omega t \Big] \\ 
m_y & \sim \delta  \cos \omega t \\
m_z & \sim \delta  \sqrt{\frac{2\beta_y + h_x}{2\beta_z + h_x}} \sin \omega t
\end{aligned}$$
where $\omega= \sqrt{(2\beta_y+h_x)(2\beta_z+h_x)}$ and the symbol $\sim$ indicates that the ratio of both sides in the equation goes to 1 as $\delta\to0$.  These solutions correspond to a trajectory with initial condition $\mb (0) = (1-\delta^2 /2 , \delta  , 0)$ and constant energy
\beq\label{energy_min}
E=-h_x +\delta^2 (\beta_y  + h_x / 2).
\eeq

\begin{figure}[htbp]
   \centering
    \includegraphics{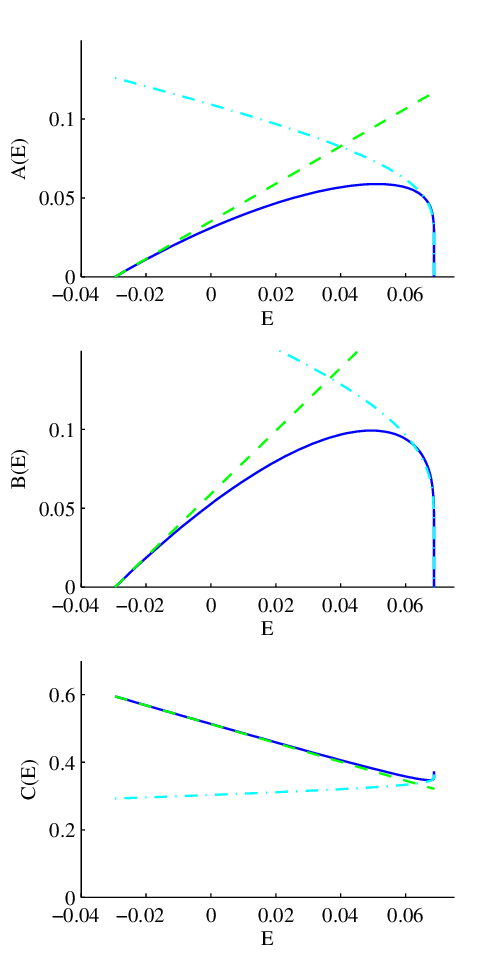}
   \caption{(color online) Comparison of the averaged energy equation coefficients on energy branch 1 computed by averaging the numerically integrated trajectories given by  $\dot{\mb}=-\mb \times \nabla E$ (blue solid line) to the asymptotic approximations in  Eq.~\eqref{minimum_averages} near the minimum energy (green dashed line) and the asymptotic approximations in Eq.~\eqref{saddle_average} near the saddle point in energy (cyan dash-dot line) for non-zero $h_x=0.03$.  The values $\beta_y=0.06$, and $\beta_z=0.5$ were used.}
   \label{fig:min_approx}
\end{figure}

To determine the averaged coefficients, we use the average defined in \eqref{avg_fnc} on the functions in the above expansion of $\mb$ over one period, $T=2\pi/\omega$ (note that $T$ does not depend on the energy in this expansion).   To write the averages in terms of the energy, we solve for $\delta^2$ as a function of $E$ from Eq.~\eqref{energy_min}, and obtain ($j=1,2$)
\beq\begin{aligned}\label{minimum_averages}
& A_j(E) \sim 2(\beta_y+\beta_z + \sigma_j h_x)(E + \sigma_j h_x)\\
&B_j(E) \sim 2(E + \sigma_j h_x)\\
&C_j(E) \sim\beta_y+\beta_z + \sigma_j h_x\\
&-\left(1+\frac{2\beta_y}{2\beta_y + \sigma_j h_x}+\frac{2\beta_z}{2\beta_z+ \sigma_j h_x}\right)(E + \sigma_j h_x) ,
\end{aligned}\eeq
where $\sigma_1=1$ and $\sigma_2=-1$.  An identical procedure was followed to obtain the expansion about $\mb_0=(-1,0,0)$ on branch 2.  The scalings in \eqref{minimum_averages} show excellent agreement to the numerically integrated averaged coefficients near the minimum energy on branch 1, as shown in Fig.~\ref{fig:min_approx}.

%
%

%
%

\subsection{Approximation Near the Saddle Point\label{sec:asym_sad}}

We proceed as in Sec.~\ref{sec:asym_min} and determine the scaling of the period of the orbit by using a series expansion of the solution. The difference is that this period goes to infinity as the energy approaches its saddle point value.  The approximate solutions are hyperbolic functions, and do not lead to complete orbits.  Rather, we can estimate the period by computing the time for the trajectory to leave a box around the saddle point, obtaining the scaling 
\beq\label{period_sad}
T(E)  = \frac{2}{\omega} \log \frac{1}{E_b - E} + O(1)
\eeq
where $\omega^2 = 4\beta_y(\beta_z-\beta_y)(1-h_x^2/4\beta_y^2)$ and $E_b = \beta_y + \frac{h_x^2}{4\beta_y}$.

To determine the approximate time-average integrals of the coefficients, we must also consider their value along the entire orbit with constant energy.  The integral is dominated by the values that $\mb$ takes along the homoclinic orbit connecting the two fixed points.  The trajectories are infinitely long and asymptotically approach the fixed points as $t \to \pm \infty$.  Therefore we take the averages to be approximated by
$ \< f(\mb (t)) \> \sim \frac{4}{T(E)} \int_0^{\infty} f(\mb (t))dt$.  We obtain the averaged energy coefficients ($j=1,2$) 
\beq\label{saddle_average}\begin{aligned}
A_j(E)&\sim \frac{4}{\nu T(E)}\left[ 4 d_j(\beta_z^2-\beta_y^2)-b_j(4\beta_y^2+h_x^2) \right] \\
B_j(E)&\sim \frac{4}{\nu T(E)}\left[ \sigma_jb_j h_x+\pi\sqrt{b_j}\beta_y\left(1-\frac{h_x^2}{4\beta_y^2}\right) \right]\\
C_j(E)&\sim \beta_z-2\beta_y+\frac{h_x^2}{4\beta_y} - \frac{4}{\nu T(E)} h_x\sqrt{b_j}\;,
\end{aligned}\eeq
where 
\begin{equation}\label{bjS}\begin{aligned}
b_j &= \left(\sqrt{S} / \beta_z +\sigma_j(\beta_z-\beta_y)h_x / 2\beta_y\beta_z \right)^2 \\
d_j & =1-\left(\sqrt{S }/ \beta_z -\sigma_j h_x / 2\beta_z \right)^2 \\
\nu^2 &= 4 S\beta_y / \beta_z + h_x(1-\beta_y/ \beta_z)\sqrt{S} ,
\end{aligned}\end{equation}
$\sigma_1=1$ and $\sigma_2=-1$, and $S=(\beta_z-\beta_y)(\beta_z-h_x^2/4\beta_y)$; details in Appendix \ref{app:saddle}.  The scalings in \eqref{saddle_average} show excellent agreement to the numerically integrated averaged coefficients near the saddle point in energy on branch 1, as shown in Fig.~\ref{fig:min_approx}.

%
\section{Results}
\label{sec:results}
%

Reducing the evolution of the magnetization vector governed by~\eqref{eq:system} to that of an energy on a graph governed by~\eqref{eq:avg_E} offers a way to better understand the effect of STT on the dynamics of the nanomagnets. At zero temperature, this approach permits to derive the full bifurcation diagram of the system: it illuminates how a new stable precessional state induced by STT is connected with a new stable fixed point in energy, as well as how STT-induced magnetization reversal is achieved by changing the stability and location of fixed points in energy.  At finite temperature, the thermally-induced switching times can be easily obtained by solving a first passage problem of the averaged equation for the energy and these times are connected to an effective energy barrier conjectured to exist.

\subsection{Bifurcation Diagram}
\label{sec:resultsA}

Here we use the reduced equation~\eqref{eq:avg_E} to obtain the bifurcation diagram of the system at zero temperature, $\epsilon=0$, and determine the fixed points of 
\begin{equation}\label{E_notemp}
\dot{E}=-\alpha A_j(E) + a_{\!J} B_j(E)
\end{equation}
and their stability.  The coefficients $A_j(E)$ and $B_j(E)$ encode the separate effects of the damping and the STT on the energy, respectively, and it can be checked that they are both zero at the critical points of the Hamiltonian  (by using the asymptotic expansions in Secs.~\ref{sec:asym_min} and \ref{sec:asym_sad} and a similar one near the energy maxima).  The energies at these points are
$$E_{a,1} = - h_x \quad \text{and}\quad E_{a,2} =  h_x$$
corresponding to the two energy minima on the lower branches 1 and 2 where $\mb =(\pm1,0,0)$, respectively;
$$E_b =\beta_y+h_x^2/4\beta_y $$
corresponding to the saddle point in energy where all four branches meet where $\mb= (-h_x/2\beta_y , \pm \sqrt{1-h_x^2 / 4\beta_y^2}, 0) $; and 
$$E_{c}= \beta_z+h_x^2/4\beta_z $$
corresponding to the two energy maxima on the upper branches 3 and 4 where $\mb= (-h_x / 2\beta_z , 0 , \pm\sqrt{1-h_x^2 / 4\beta_z^2})$. These critical points can merge and disappear when the applied field crosses the critical values $h_x = \pm 2 \beta_y$ and $h_x = \pm 2 \beta_z$.  In addition, only the two energy minima can ever be stable, and one of them looses stability when another nontrivial fixed point in energy, $E_0$, appears on one of the energy branches.

The non-trivial fixed point of \eqref{E_notemp} appears for certain values of $\aJ$ at energy $E_0$, and corresponds to the stable precessional state.  At $E_0$, the energy lost by damping, $-\alpha A_j(E_0)$, is exactly compensated by the energy gained by STT, $\aJ B_j(E_0)$:
\begin{equation}
\label{eq:E0}
-\alpha A_j(E_0)+ a_{\!J} B_j(E_0) =0.
\end{equation} 
The stable fixed point at $E_0$ does not corresponds to a stable fixed point of the original dynamics at finite $\alpha$, but rather to a stable limit cycle (precessional state), see Fig.~\ref{fig:orbits} for a schematic illustration.  Interestingly, in  the case of uniaxial anisotropy (see Appendix B), the new fixed point is always unstable, therefore no precessional state is seen.

\begin{figure}
   \centering
   \includegraphics[scale=0.9]{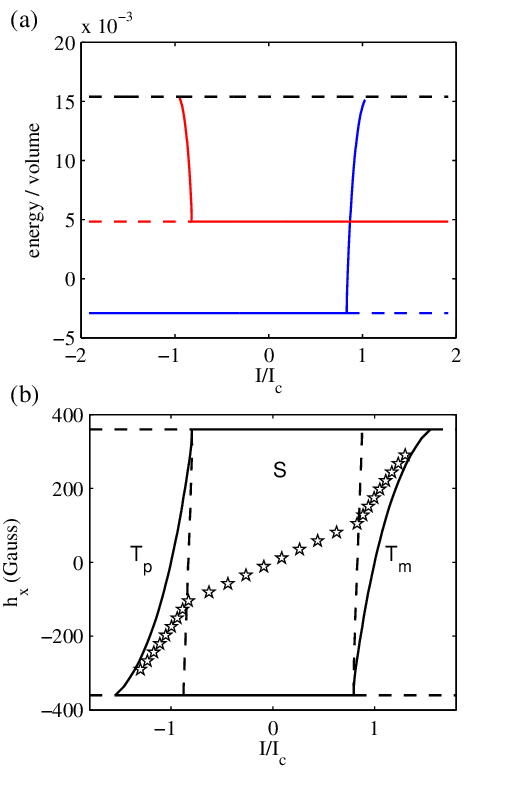}
   \caption{(color online) (a): Bifurcation diagram as a function of current $I$ (scaled by the critical current $I_c$ to induce switching at $h_x=0$) at the fixed value of $h_x=34.8$ G: the stable (solid line) and unstable (dashed line) fixed points of $\dot{E}=-\alpha A_j(E)+a_{\!J} B_j(E)$ are shown for both energy branch 1 (blue) and 2 (red).  (The remaining two unstable fixed point with higher energy are not shown.)  (b): Bifurcation diagram as a function of $I/I_c$ and $h_x$ (compare with experimental data shown in Fig.~2a in Ref.~[\onlinecite{KEGSKRB2004}]).  The dashed lines on either side of region $S$ correspond to the current required to first initiate a stable precessional state ($I$ from $a_{\!J}$ given by~\eqref{eq:ac_p}) while the solid lines correspond to the current required to induce switching ($I$ from $a_{\!J}$ given by~\eqref{eq:ac}).  Beyond the solid lines, only one stable fixed point remains: in region $T_p$ ($T_m$) it is $m_x=+1$ ($m_x=-1$) and for $h_x$ beyond the region shown, only one lower energy branch remains.  The stars indicate where the mean thermally induced switching times from branch 1 to 2 and branch 2 to 1, computed via~\eqref{ODE_tau}, are equal. }
   \label{fig:phase}
\end{figure}

From the location and the stability of the fixed points identified above, shown in Fig.~\ref{fig:phase}(a) as a function of  current, $I$, for a fixed value of~$h_x$, we can understand how magnetization reversal is achieved by varying the strength of the spin-polarized current:  A positive current destabilizes the minimum at $(1,0,0)$ on branch $j=1$ and eventually the fixed point at $E_0$ is also lost, leaving the only stable fixed point at $(-1,0,0)$ on branch $j=2$.  As the coefficient $B_2(E)$ has the opposite sign of $B_1(E)$ while $A_2(E)$ and $A_1(E)$ are always positive, negative current is required to switch the magnetization back again.

We can also calculate the full bifurcation diagram shown in Fig.~\ref{fig:phase}(b), which is remarkably similar to the experimental one (see Fig.~2a in Ref.~[\onlinecite{KEGSKRB2004}]).  One of the energy minima looses its stability and the precessional state appears when $E_{a,1}$ or $E_{a,2}$ solves~\eqref{eq:E0}, i.e. when $a_{\!J}$ is given by ($j=1,2$)
\begin{equation}\label{eq:ac_p}
  a_{\!J}= \alpha \lim_{x\downarrow E_{a,j}} \frac{A_j(x)}{B_j(x)} = \alpha\sigma_j (\beta_y+\beta_z +\sigma_j h_x), 
\end{equation}
where $\sigma_1=1$ and $\sigma_2=-1$.  
The corresponding boundaries on the bifurcation diagram are shown as dashed lines in Fig.~\ref{fig:phase}(b).  The limit in \eqref{eq:ac_p} was obtained using asymptotic expansions of the coefficients in \eqref{minimum_averages}. The precessional state exists in the region between the dashed and the solid lines in Fig.~\ref{fig:phase}(b). Beyond these solid lines only one stable state remains. This occurs when $E_b$ solves~\eqref{eq:E0}, meaning that the strength of the current required to induce switching is ($j=1,2$)
 \begin{equation}
   \label{eq:ac}
   \begin{aligned}
     a_{\!J} &= \alpha \lim_{x\uparrow E_{b}}
     \frac{A_j(x)}{B_j(x)}\equiv \lambda_j\\
     &=\alpha\sigma_j \frac{4
       d_j(\beta_z^2-\beta_y^2)-b_j(4\beta_y^2+h_x^2)}{\sigma_j
       b_j h_x+\pi\sqrt{b_j}\beta_y(1-h_x^2/4\beta_y^2)}
   \end{aligned}  
\end{equation}
where $b_j$, $d_j$, and $S$ are define in \eqref{bjS}.   This reduced to $\lambda_j = \alpha\sigma_j 4\sqrt{\beta_z(\beta_z-\beta_y)}/\pi$ when $h_x=0$.  The limit was taken using the expansion of the coefficients in \eqref{saddle_average}.  Note that the dimensional electric current, $I$, in  \eqref{eq:aJ} is simply a scaled version of $\aJ$, therefore we present our results in terms of $I/I_c = \aJ/\lambda_1$, where $\lambda_1$ is computed for $h_x=0$. 

In contrast, for the case of uniaxial anisotropy, the new unstable fixed point appears immediately for non-zero values of $\aJ$, there is no stable precessional state, and the critical current to induce switching is given by \eqref{uni_crit} in Appendix~\ref{sec:uni},
$$ \aJ  = \alpha\sigma_j(  2\beta_x +\sigma_j h_x) $$
for energy of the form $E=\beta_x(1-m_x^2) - h_xm_x$.

\subsection{Thermally-Induced Transitions} 
\label{sec:resultsB}

\begin{figure}
   \centering
  \includegraphics{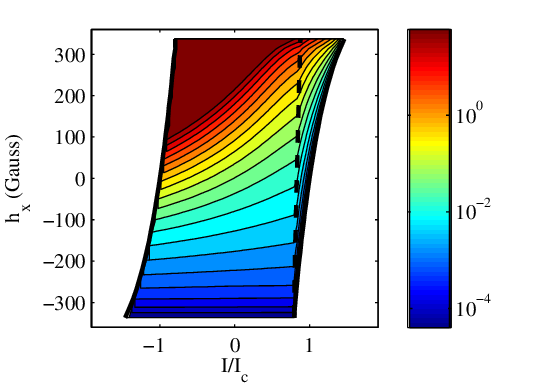}
   \caption{(color online) Contour plot of the mean first passage time (in seconds) from the fixed point on energy branch $j=1$, to the fixed point on branch 2.  (The times from branch 2 to 1 would be the same figure, rotated $180^\circ$.) These times are computed using~(\ref{ODE_tau}) and plotted as a function of $h_x$ and current $I$ normalized by the current $I_c$ to induce switching at $h_x=0$. The shape of the plotted region is identical to that shown in Fig.~\ref{fig:phase}(b). Between the dashed line ($I$ from $a_{\!J}$ in~\eqref{eq:ac_p}) and solid black line ($I$ from $a_{\!J}$ in~\eqref{eq:ac}), the starting point is $E_0$ (Eq.~\eqref{eq:E0}); everywhere else it is the minimum energy, $E_{a,1}$.  Switching times greater than one minute are all colored in dark red.  }
   \label{fig:tau}
\end{figure}

Here, we study thermally induced magnetization reversal. To this end we use the reduced system in~(\ref{eq:avg_E}) with $\epsilon>0$ to calculate the mean transition times (i.e.~dwell times) between the stable fixed points of the deterministic dynamics identified before.  The mean time~$\tau_j(x)$ to transition from energy $x$ on branch $j=1,2$ to the fixed point on the other branch satisfies
\begin{equation}
  \label{ODE_tau}
  \begin{aligned}
    \big[-\alpha A_j(x) +  a_{\!J}B_j(x) +&2\alpha\epsilon C_j(x)\big]\tau_j'(x) \\
    & +  \alpha\epsilon A_j(x)\tau_j''(x) = -1
  \end{aligned}
\end{equation}
with a matching condition (see Appendix~\ref{sec:Matching}) to prescribe transitions through the center node of the graphs shown in Figs.~\ref{fig:orbits}(b), (c) as well as an absorbing condition at the target state.  Equation \eqref{ODE_tau} is valid at any temperature and its solution can be expressed in terms of integrals involving the coefficients $A_j(x)$, etc. Evaluating these integrals numerically leads to the results shown in Fig.~\ref{fig:tau}. We can also evaluate these integrals asymptotically in the limit when the temperature is small ($\epsilon\ll1$), in which case they are dominated by the known behavior of the coefficients near the critical points. These calculations are tedious but straightforward and reported in Appendix~\ref{sec:MFPT}.  In situations where the system transits from the stable minimum $E_{a,1}$ or $E_{a,2}$ on branch 1 or 2 to the stable point (minimum $E_{a,2}$ or $E_{a,1}$ or precessional state $E_0$) on the other branch we obtain ($j=1,2$)
\beq\label{tau_min}
\tau_j \sim \frac{\gamma_1+\gamma_2}{\gamma_j} \frac{\epsilon}{\alpha}\frac{e^{(1-\frac{\aJ}{\lambda_j}) (E_b-E_{a,j})/\epsilon}}{2(\beta_y+\beta_z \pm h_x)(1- \frac{\aJ}{\lambda_j} )^2}  
\eeq
whereas in situations when switching occurs from the stable precessional state $E_0$ we obtain ($j=1,2$ depending on whether $E_0$ is on branch 1 or 2)
\beq\label{tau_E0}
\tau_j \sim \frac{\gamma_1+\gamma_2}{\gamma_j} \frac{\epsilon}{\alpha}\frac{e^{(1-\frac{\aJ}{\lambda_j})(E_b-E_0)/\epsilon}}{A_j(E_0)(1- \frac{\aJ}{\lambda_j} )^2}  .
\eeq
Here $\lambda_j$, defined in~\eqref{eq:ac}, is the critical current to induce switching at zero temperature (which depends on $h_x$), and
$$ \begin{aligned}
\gamma_1 = 4d_2(\beta_z^2-\beta_y^2)-b_2(4\beta_y^2+h_x^2)\\
\gamma_2 = 4d_1(\beta_z^2-\beta_y^2)-b_1(4\beta_y^2+h_x^2)
\end{aligned}$$
with $b_j$ and $d_j$ defined in~\eqref{bjS}. 

The results in~\eqref{tau_min} and \eqref{tau_E0} agree with the experimental observations~\cite{KEGSKRB2004,MASBBR2002} and the theoretical predictions~\cite{AV2005,LZ2004} that the effect of STT on the dwell times can be captured via a N\'eel-Brown-type formula with an effective energy scaling linearly with the current strength. We stress, however, that these previous theoretical works had to assume the existence of such a formula, whereas \eqref{tau_min} and \eqref{tau_E0} fall out naturally from the asymptotic analysis, and give explicit expressions not only for the effective energy but also the prefactors and their dependency on the strength of the current producing STT.

\section{Conclusions\label{sec:conclusions}}

In summary, we have shown how the dynamical behavior of nanomagnets driven by spin-polarized currents can be understood in the low-damping regime by mapping their evolution to the diffusion of an energy on a graph. We thereby obtained the full bifurcation diagram of the magnet at zero-temperature as well as the mean times of thermally assisted magnetization reversal. These results agree with experimental observations and give explicit expressions for the dwell times in terms of a N\'eel-Brown-type formula with an effective energy, thereby settling the issue of the existence of such a formula. 

We carried the analysis for micromagnets that are of the specific type considered by Li and Zhang~\cite{LZ2003}, but the method presented in this paper is general and can be applied to other situations with different geometry, applied fields that are time-dependent or not, etc. Without any additional computations, the STT current $\aJ$ in \eqref{eq:avg_E} could be made time varying to understand the effect of pulse width or the magnetization reversal time when the current is switched on.  Investigating the effect of the direction of the STT current, ${\bf m}_p$, only requires recomputing the coefficient $B_j(E)$ for the new direction.  If the form of the energy in \eqref{eq:3} were to be changed, then the coefficients in \eqref{eq:coeff} would change, and the asymptotic analysis of the coefficients would need to be repeated for the new Hamiltonian system (see Appendix B for one such case).  Our averaging method can also be applied to systems in which the magnetization varies spatially in the sample.  In these situations, the graph of the energy will be more complicated, but the general procedure to reduce the dynamics to a diffusion on this graph remains the same. Such a study will be the object of a future publication.

\begin{acknowledgments} We would like to thank Dan Stein and Andy Kent for useful discussions.  The research of E.~V.-E. was supported in part by NSF grant DMS07-08140 and ONR grant N00014-11-1-0345.  \end{acknowledgments}

\appendix

\section{Converting to Ito Equation for Energy\label{app:Ito}}

In this Appendix, we explicitly show the steps of converting \eqref{eq:system} to an Ito SDE and determining $\dot{E}$ using the stochastic chain rule, in preparation for obtaining Eq.~\eqref{eq:avg_E} in the text.  First, we write the Strotonovich SDE~\eqref{eq:system} in the form
\beq\label{eq:system_ito}
 \frac{d\mb}{dt} = {\bf a}_c -\alpha {\bf a}_d + a_J {\bf a}_p + \sqrt{\frac{2\alpha\epsilon}{1+\alpha^2}} \mathcal{B} \etab(t)
\eeq
where the conservative term is
$$
{\bf a}_c = \left( \begin{array}{c}
2(\beta_z-\beta_y)m_ym_z \\ -m_z(2\beta_z m_x+h_x) \\ m_y(2\beta_y m_x + h_x) 
\end{array}\right)
$$
the damping term is
$$
{\bf a}_d = \left( \begin{array}{c}
-h_x(1-m_x^2)-2m_x(\beta_ym_y^2+\beta_zm_z^2) \\
m_y(h_xm_x+2\beta_y(1-m_y^2)-2\beta_zm_z^2) \\
m_z(h_xm_x-2\beta_ym_y^2 + 2\beta_z(1-m_z^2)) 
\end{array}\right)
$$
the spin-torque transfer term is
$$
{\bf a}_p = \left( \begin{array}{c}
m_x^2 - 1\\ m_xm_y \\ m_xm_z
\end{array}\right)
$$
and the diffusion matrix, $ \mathcal{B} $ equals  
$$
\left( \begin{array}{ccc}
 \alpha(1-m_x^2) & m_z - \alpha m_x m_y & -m_y - \alpha m_x m_z\\
-m_z-\alpha m_xm_y & \alpha( 1-m_y^2) & m_x-\alpha m_ym_z\\
m_y-\alpha m_xm_z & -m_x - \alpha m_ym_z & \alpha(1-m_z^2)
 \end{array}\right).
$$
In order to convert this to its Ito form, the drift term obtains the correction 
$${\bf a}_I =\frac{2\alpha\epsilon}{1+\alpha^2} \frac{1}{2} \sum_{j,k}\mathcal{B}_{kj}\partial_k \mathcal{B}_{ij} = - 2\alpha\epsilon {\mb},$$
making the Ito SDE for the magnetization direction
\beq \label{Ito_m}
\frac{d\mb}{dt} = {\bf a}_c -\alpha {\bf a}_d + a_J {\bf a}_p -2\alpha\epsilon{\mb} + \sqrt{\frac{2\alpha\epsilon}{1+\alpha^2}} \mathcal{B} \etab (t).
\eeq

We use the rules of Ito calculus to compute  $\dot E = \nabla_m E \cdot \dot{\mb}$ and obtain
\beq \label{E_noavg}
\dot{E}=-\alpha A(\mb)+a_J B(\mb) + 2\alpha\epsilon C(\mb)+\sqrt{2\alpha\epsilon } \sqrt{A(\mb)}\xi(t),
 \eeq
where $\xi(t)$ is 1D white noise.  Notice since the term ${\bf a}_c$ in Eq.~\eqref{Ito_m} conserves energy, it has no corresponding term in Eq.~\eqref{E_noavg}.  The remaining terms in Eq.~\eqref{Ito_m} have corresponding terms in Eq.~\eqref{E_noavg}:  the dissipative term, ${\bf a}_d$, leads to
$$
A(\mb) = 4(\beta_y^2m_y^2+\beta_z^2m_z^2 - E^2) - 4Eh_xm_x  +  h_x^2(1-m_x^2),
$$
the spin-torque transfer terms, ${\bf a}_p$, leads to
$$ 
B(\mb) =  2Em_x + h_x(1+m_x^2),
$$
and the correction term for Ito calculus ($\partial_j$ indicates partial derivative with respect to the $j$th element of $\mb$),
$$\begin{aligned}
\frac{2\alpha\epsilon}{1+\alpha^2}\frac{1}{2}\sum_{i,j} & [\mathcal{B}\mathcal{B}^T]_{ij}\partial_{i}\partial_{j} E(\mb) \\
& = 2\alpha\epsilon \Big(\beta_y(1-m_y^2)+\beta_z(1-m_z^2)\Big),
\end{aligned}$$
together with the contribution from ${\bf a}_I$ gives
$$
C(\mb) =   \beta_y + \beta_z  - 3E - 2h_xm_x .
$$
The strength of the noise term in \eqref{E_noavg} is computed from the combination of the three strengths 
$$\begin{aligned}
b_x(\mb) &= 2 (\beta_z - \beta_y) m_y m_z \\ 
& - \alpha [h_x (1-m_x^2) + 2m_x (\beta_y m_y^2 + \beta_z m_z^2)]\\
b_y(\mb) &= - m_z( h_x  + 2 \beta_z m_x) \\
& + \alpha [h_x m_x m_y + 2m_y (\beta_y (1 - m_y^2) - \beta_z m_z^2)] \\
b_z(\mb) &= h_x m_y + 2 \beta_y m_x m_y \\
& + \alpha [h_x m_x m_z + 2 m_z ( \beta_z (1 - m_z^2)- \beta_y m_y^2 )],
\end{aligned}$$
of the three independent components of the white noise, $\etab(t)$, in Eq.~\eqref{Ito_m}.  These simplify to $\sqrt{b_x^2+b_y^2+b_z^2 }=\sqrt{(1+\alpha^2)A(\mb)}$, giving the noise amplitude in Eq.~\eqref{E_noavg}.   

%
%
\section{Magnet with Uniaxial Anisotropy\label{sec:uni}}

Here we highlight the difference between the magnet with biaxial anisotropy discussed in the main text and one with uniaxial anisotropy; its energy is given by
$$ E = \beta_x(1-m_x^2) - h_xm_x .$$
The system of equations for $\mb$ are equivalent to system \eqref{eq:system_ito} but with $\beta_y$ and $\beta_z$ replaced by $\beta_x$.  The averaged evolution of the energy, obtained from $\dot{E}=\nabla_m E \cdot \dot{\mb}$, is 
\begin{equation}
\dot{E} = -\alpha A_j(E) + \aJ B_j(E) + 2\alpha\epsilon C_j(E) + \sqrt{2\alpha\epsilon A_j(E)}\xi(t)
\end{equation}
for $j=1,2$ where
$$\begin{aligned}
A_j(E) &= (1-\<m_x^2\>_j)(2\beta_x\<m_x\>_j+h_x)^2 \\
B_j(E) &= (1-\<m_x^2\>_j)(2\beta_x\<m_x\>_j+h_x) \\
C_j(E) &= 3\beta_x\<m_x^2\>_j +h_x\<m_x\>_j - \beta_x 
\end{aligned}$$
and $\xi(t)$ is standard one-dimensional white noise.  The Hamiltonian orbits in this case reduce to circles at constant $m_x$, producing a direct relationship between $m_x$ and the energy $E$ (i.e.~$\<m_x\>_j = m_{x,j}$) given by
\begin{equation}
m_{x,j} = \frac{-h_x}{2\beta_x} +\sigma_j \sqrt{\left(1-\frac{E}{\beta_x}\right) + \frac{h_x^2}{4\beta_x^2} }.
\end{equation}
The graph is only two branches, with $\sigma_1=1$ on the one corresponding to $m_x>-h_x/2\beta_x$ and $\sigma_2=-1$ on the other with $m_x<-h_x/2\beta_x$.  They meet at the saddle point energy, $ E_b= \beta_x + h_x^2/4\beta_x$ where there is an entire circle of critical points given by $m_x=-h_x/2\beta_x$.

\begin{figure}
   \centering
   \includegraphics{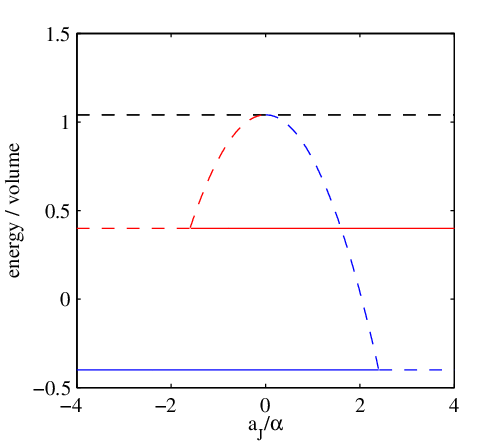} 
   \caption{Bifurcation Diagram: location and stability of fixed points in the uniaxial anisotropy case, with $\beta_x=1$ and fixed $h_x=0.4$ for illustrative purposes.  The minimum on branch 1 (blue) and branch 2 (red) loose stability when the new fixed point given in \eqref{eq:E0} merges with it.}
   \label{fig:E0}
\end{figure}

As with the biaxial case, we can find the critical current to induce switching by studying the bifurcation diagram of the deterministic system, $\dot{E}=-\alpha A_j(E) + \aJ B_j(E)$.  The fixed point not corresponding to one of the critical values of energy has energy 
\begin{equation}\label{eq:E0_uni}
E_0 = \left(\beta_x+\frac{h_x^2}{4\beta_x}\right) - \left(\frac{\aJ}{\alpha}\right)^2\frac{1}{4\beta_x}
\end{equation}
when this $E_0$ solves $\alpha A_j(E) = \aJ B_j(E)$.  Note while the value of $E_0$ in \eqref{eq:E0_uni} is the same for both $\pm \aJ$, only one of these corresponds to a fixed point.  From the bifurcation diagram in Fig.~\ref{fig:E0} we immediately see differences from the biaxial case: the new fixed point immediately emanates from the saddle fixed point, is unstable (there is no stable precessional state), and the minimum looses stability only after the new fixed point merges with it.  Thus, we find the critical current 
\begin{equation}\label{uni_crit}
\aJ = \alpha\sigma_j(  2\beta_x +\sigma_j h_x)
\end{equation}
to induce switching is when $E_0$ equals the minimum energy, $E_{a,j}=\sigma_j h_x$,
with $\sigma_1=1$ for the $m_x=1$ to $m_x=-1$ switch and $\sigma_2=-1$ for the $m_x=-1$ to $m_x=1$ switch.  Notice that $\aJ$ in \eqref{uni_crit} is the same as \eqref{eq:ac_p} with $\beta_y=\beta_z=\beta_x$, but this critical value for the biaxial case is for the appearance of the stable precessional state and not the critical current to induce switching, which is given by the value of $\aJ$ in \eqref{eq:ac}.  As $\beta_y\to\beta_z$ the value of $\aJ$ in  \eqref{eq:ac} becomes less than the value of $\aJ$ in \eqref{eq:ac_p}, thus changing the stability of the new fixed point from stable to unstable. The critical value of $\aJ$ to induce switching is now given by \eqref{eq:ac_p}.  Once $\beta_y=\beta_z$ the value of $\aJ$ in  \eqref{eq:ac} has become zero; the new fixed point immediately emerges from the saddle point energy.

%
%

\section{Integration Near Homoclinic Orbit\label{app:saddle}}

Here, we compute the integral $\int_0^{T(E)} f(\mb (t)) dt$, which is dominated by the dynamics of $\mb$ on the homoclinic orbit with energy $E_b$, in order to obtain the scalings of the averaged coefficients in \eqref{saddle_average}.  We use an  approximate  trajectory of $\mb$ that starts at $t=0$ at the point on the orbit midway between the two fixed points, where $m_z$ is positive and $m_y=0$.  The trajectories are infinitely long and asymptotically approach the fixed points.  Therefore we approximate
$$ \int_0^{T(E)} f(\mb (t)) dt \sim 4\int_0^{\infty} f(\mb (t))dt$$ 
and we take the averages to be approximated by
\beq\label{approx_avg}
\< f(\mb (t)) \> \sim \frac{4}{T(E)} \int_0^{\infty} f(\mb (t))dt.
\eeq

For the simple case when $h_x=0$, the exact solution to the Hamiltonian system is
\beq\label{m_h0}\begin{aligned}
m_x(t) & = \sqrt{\frac{\beta_z-\beta_y}{\beta_z}}\textrm{sech}\nu_0 t\\
 m_y(t) &= \tanh \nu_0 t \\
 m_z(t) & = \sqrt{\frac{\beta_y}{\beta_z}} \textrm{sech} \nu_0 t 
\end{aligned} \eeq
where $\nu_0=2\sqrt{\beta_y(\beta_z-\beta_y)}$. For non-zero $h_x$, an exact solution is unknown, but the components $m_x$ and $m_z$ are well approximated by sech functions, with appropriate values to match the exact solution at $t=0$ and as $t\to\pm\infty$.  These are
\beq\label{m_h}\begin{aligned}
m_x&\sim   \frac{-h_x}{2\beta_y}+A_x\textrm{sech} \nu t =  \\
& \frac{-h_x}{2\beta_y}  + \left[ \sqrt{\frac{\beta_z-\beta_y}{\beta_z}\Big(1-\frac{h_x^2}{4\beta_y\beta_z}\Big)} + \frac{\beta_z-\beta_y}{2\beta_y\beta_z}h_x \right] \textrm{sech}\nu t \\
m_z&\sim A_z\textrm{sech}\nu t \\
& =  \sqrt{1-\Big(\sqrt{\frac{\beta_z-\beta_y}{\beta_z}\left(1-\frac{h_x^2}{4\beta_y\beta_z}\right)}-\frac{h_x}{2\beta_z} \Big)^2 }\;\textrm{sech}\nu t
\end{aligned}\eeq
where 
$$\begin{aligned}
\nu^2 & = 4(\beta_y-h_x^2/4\beta_z)(\beta_z-\beta_y) \\
& + h_x\frac{\beta_z-\beta_y}{\beta_z}\sqrt{4(\beta_z-\beta_y)(\beta_z-\frac{h_x^2}{4\beta_y})}
\end{aligned}$$
was found by matching the second derivative at $t=0$ to the second derivative found from the Hamiltonian system.  The coefficient $A_x$ was found by noting that $m_y(0)=0$ and then solving $\beta_y+h_x^2/(4\beta_y) = \beta_z(1- m_x^2(0)) - h_xm_x(0)$ for $A_x$.  Then, the coefficient $A_z=\sqrt{1-m_x^2(0)}$.  These solutions are also consistent with $m_x(t)\to \frac{-h_x}{2\beta_y}$ and $m_z(t)\to 0 $ as $t\to\infty$.  Furthermore, the solutions in \eqref{m_h} reduce to the above exact solutions in \eqref{m_h0} when $h_x=0$.  

For computing the averages, we first note that
$$\begin{aligned}
\<m_x\>& =\frac{-h_x}{2\beta_y} + A_x\< \textrm{sech}\nu t \> \\
\< m_x^2\> & = \frac{h_x^2}{4\beta_y^2} - \frac{h_x}{\beta_y}A_x\< \textrm{sech}\nu t \> + A_x^2  \< \textrm{sech}^2\nu t \>\\
\end{aligned}$$
and 
$$ \< m_y^2\> = 1 - \<m_x^2\> - \< m_z^2\> $$
using the constraint that the magnetization vector has unit length.  After computing the approximate average defined in \eqref{approx_avg}, we obtain
$$\begin{aligned}
\<m_x\>& \sim   \frac{-h_x}{2\beta_y} + \frac{2\pi A_x}{\nu T(E)} \\
\<m_x^2\> & \sim \frac{h_x^2}{4\beta_y^2} - \frac{h_x2\pi A_x}{\beta_y\nu T(E)}+ \frac{4A_x^2}{\nu T(E)}\\
\< m_z^2\> & \sim \frac{4A_z^2}{\nu T(E)}\\
\<m_y^2\>&\sim 1 -  \frac{h_x^2}{4\beta_y^2} +\left( \frac{\pi h_xA_x}{2\beta_y} - A_x^2 - A_z^2\right) \frac{4}{\nu T(E)} 
\end{aligned}$$
where $T(E)\sim -\frac{2}{\omega}\log (E_b-E)$ when $E_b=\beta_y+ h_x^2/4\beta_y$ and $\omega= 2\sqrt{(\beta_y-h_x^2/4\beta_y)(\beta_z-\beta_y)}$.  The averaged coefficients in \eqref{saddle_average} follow.

%
\section{Matching Conditions\label{sec:Matching}}
%

In this Appendix, we derive the matching conditions for the mean first passage time equation \eqref{ODE_tau}.  Matching conditions are required only at the saddle point where the energy branches meet~\cite{FWbook} because it is a regular boundary point (see~\cite{Feller1954} for boundary point classification); it is accessible from the interior of each energy branch and the interior of each energy branch is accessible from it.  On the other hand, no additional boundary conditions are required at the other ends of the energy branches, specifically the minima, as these are entrance boundary points; the interior of the energy branches are accessible from these points, but the expected passage time from the interior to these points is infinite.  This coincides with the diffusion of the magnetization vector on the surface of the unit sphere.  The original SDE for the magnetization vector contains no extra conditions prescribed at the single points corresponding to the energy minima and maxima.  

From the matching conditions, we are able to construct the probabilities of the energy switching from one branch to another (the matching conditions required to supplement Eq.~(\ref{eq:avg_E}) in the text) as well as the pre-factors for the mean first passage times in Eqs.~(\ref{tau_min}) and (\ref{tau_E0}) in the text describing the probability the system switches to the other lower energy branch rather than return to the original one.  The derivation is based on the conservation of probability flux of the magnetization vector across the homoclinic orbit on the sphere with energy equal to the saddle point energy, $E_b$.  For ease of notation, any function evaluated at energy $E_b$ should be interpreted as a limit as $E\to E_b$ from the interior of the energy branch.  

Consider $\rho_j(E,t)$ to be the probability density for the energy while on branch $j$, normalized so that
$$ \begin{aligned}
\int_{E_{a,1}}^{E_b} &\rho_1(E,t) dE +  \int_{E_{a,2}}^{E_b} \rho_2(E,t) dE +  \int_{E_{b}}^{E_c} \rho_3(E,t) dE \\
& +  \int_{E_{b}}^{E_c} \rho_4(E,t) dE = 1.
\end{aligned} $$ 
These density functions are continuous at the saddle point in energy:
$$
\rho_1(E_b,t) = \rho_2(E_b,t) = \rho_3(E_b,t) = \rho_4(E_b,t).
$$
The functions $A_j(\mb(t))$ are also continuous across the homoclinic orbit, therefore, if this orbit is approached from the higher or the lower energy branches, we have that
$$\begin{aligned}
 \int_0^{T_1(E_b)} & A_1(\mb(t)) dt + \int_0^{T_2(E_b)} A_2(\mb(t)) dt \\
 & =
 \int_0^{T_3(E_b)} A_3(\mb(t)) dt +  \int_0^{T_4(E_b)} A_4(\mb(t)) dt
\end{aligned}$$
or equivalently
\beq\label{cons_A}\begin{aligned}
 & T_1(E_b) A_1(E_b) +  T_2(E_b) A_2(E_b) \\
 & =
 T_3(E_b) A_3(E_b)  + T_4(E_b) A_4(E_b) .
\end{aligned}\eeq
This provides the understanding for why the flux of the total probability,  $ T_j(E)\rho_j(E,t)$, and not simply the averaged probability, $\rho_j(E,t)$, is conserved across the homoclinic orbit.

The forward Kolmogorov equation for the total probability density, written in terms of the flux, $J_j[\cdot]$, on each branch $j=1,2,3,4$, is
\beq\label{FP}
\frac{\partial}{\partial t} T_j(E)\rho_j(E,t) = - \frac{\partial}{\partial E} J_j[T_j(E) \rho_j(E,t)]
\eeq
where
$$\begin{aligned}
J_j&[T_j(E)\rho_j(E,t)] = - \alpha\epsilon \frac{\partial}{\partial E} \Big( A_j(E)T_j(E)\rho_j(E,t) \Big) \\
& + \Big[-\alpha A_j(E) + \aJ B_j(E) + 2\alpha\epsilon C_j(E)\Big] T_j(E)\rho_j(E,t) .
\end{aligned}$$
Analogous to Eq.~\eqref{cons_A}, the conservation of probability flux across the homoclinic orbit provides the matching condition for Eq.~\eqref{FP}:
\beq\label{Pflux}\begin{aligned}
J_1&[T_1(E_b) \rho_1(E_b,t)] + J_2[T_2(E_b) \rho_2(E_b,t)] \\
&= J_3[T_3(E_b) \rho_3(E_b,t)] + J_4[T_4(E_b) \rho_4(E_b,t)].
\end{aligned}\eeq
The differential equation~(\ref{ODE_tau}) in the main text for the mean exit time, $\tau_j(E)$,  from energy $E$, comes from the backwards Kolmogorov equation; it uses the adjoint operator to the one in \eqref{FP}.  Therefore, Eq.~(\ref{ODE_tau})'s matching condition is the adjoint condition to the conservation of probability flux, Eq.~\eqref{Pflux}.  After dividing by $\alpha\epsilon$, the matching condition for Eq.~(\ref{ODE_tau}) in the text is
\beq\begin{aligned}\label{matching}
A_1&(E_b)T_1(E_b) \tau_1'(E_b) + A_2(E_b)T_2(E_b) \tau_2'(E_b) = \\
& A_3(E_b)T_3(E_b)\tau_3'(E_b) + A_4(E_b)T_4(E_b)  \tau_4'(E_b).
\end{aligned}\eeq
This condition is equivalent to the condition stated by Fredlein and Wetzell~\cite{FWbook}.


From the exit time matching condition, \eqref{matching}, we derive the probabilities for the energy to switch branches in order to complete the stochastic differential equation (\ref{eq:avg_E}) describing the evolutions of the energy, as well as determine the pre-factor for the mean switching times between meta-stable states appearing in Eqs.~(\ref{tau_min}) and (\ref{tau_E0}).  

We define the notation $P(j \to k)$ to be the probability the energy switches from energy branch $j$ to energy branch $k$ at the saddle point, $E_b$.  In general, this probability is derived from the coefficients of the matching condition~\eqref{matching} by breaking the integral within the coefficients $A_j(E_b)$ into the parts which lead to each of the other energy branches; these fractional parts out of the whole integral yield the probabilities $P(j\to k)$.  Further simplifications are made by taking advantage of the symmetry of this particular magnetic system.

In general, the probability, $P(j \to k)$, to switch from branch $j$ to branch $k$ is
$$
P(j \to k) = \frac{ \int_0^{T_j(E_b)} A_j(\mb(t)) 1_{k}(\mb(t)) dt}{  \int_0^{T_j(E_b)} A_j(\mb(t)) dt}
$$
where $1_{k}(\mb(t)) = 1$ if $\mb(t)$ is closer to orbits in branch $k$ than any of the other branches besides the one in which it resides, and 0 otherwise.  Immediately from Fig.~1(a) in the text, we see that
\begin{subequations}\label{erg_matching}
\beq\label{P12}
 P(1 \to 2) = P(2 \to 1) = P(3 \to 4) = P(4 \to 3) = 0
 \eeq
since $1_{k}(\mb(t)) = 1$ at only two individual points (at the green dots).  Exploiting the symmetry about the $m_x$-$m_y$ plane, we have that
\beq
P(1 \to 3 ) = P(1 \to 4) = 1/2, 
\eeq
\beq
P(2 \to 3 ) = P(2 \to 4) = 1/2, 
\eeq
\beq
 P(4 \to 1) =  P(3\to 1), 
 \eeq
and 
\beq
 P(4 \to 2) =  P(3 \to 2) .
 \eeq
Using the above, we can rewrite
\beq\label{P31}\begin{aligned}
P(3\to 1)  = \frac{g_1}{g_1+g_2}
\end{aligned}\eeq
where we define the notation
$$
g_j =  \int_0^{T_j(E_b)} A_j(\mb(t)) dt   
$$
 for $j=1,2$.  We can take 
 $$
g_j \approx 4 d_j (\beta_z^2-\beta_y^2)-b_j(4\beta_y^2+h_x^2),
$$
coming from Sec.~\ref{sec:asym_sad} with out the term $4/\nu$, since $g_j$ only appears as fractions.  Similarly to \eqref{P31}, we have that
\beq
 P(3\to 2) = \frac{g_2}{g_1+g_2}.
\eeq
\end{subequations}
All together, the conditions \eqref{erg_matching} provide the switching probabilities for the stochastic energy equation (\ref{eq:avg_E}) in the text.

The mean first passage time calculation requires the probability the energy switches from branch 1 to 2 or 2 to 1, which we can see from Eq.~\eqref{P12} never happens along a direct path.  Rather, the energy must first switch to one of the two higher energy branches.  Conditioning on which intermediate branch the energy switches to, we have that the probability the energy switches from branch 1 to branch 2 is
$$\begin{aligned}
P( \textrm{switch from }1 )  =& P(1 \to 3)P(3 \to 2) \\ & + P(1 \to 4) P(4 \to 2).
\end{aligned}$$
Using the simplified probabilities in \eqref{erg_matching}, we have that 
$$
P( \textrm{switch from }1 ) = \frac{ P(3 \to 2)}{2}  + \frac{P(3 \to 2)}{2} = \frac{g_2}{g_1+g_2}.
$$
Similarly, the switching from energy branch 2 back to 1 is
 $$
 P( \textrm{switch from }2 ) =  \frac{g_1}{g_1+g_2}.
 $$
To match the notation in the text, we define the switching probability from branch $j$ to be
\beq\label{Pswitch}
P( \textrm{switch from }j ) = \frac{\gamma_j}{\gamma_1+\gamma_2}
\eeq
where $\gamma_1=g_2$ and $\gamma_2=g_1$.  The probabilities in \eqref{Pswitch} are precisely the pre-factors in Eqs.~(\ref{tau_min}) and (\ref{tau_E0}) for the mean switching times.

\section{Mean First Passage Time\label{sec:MFPT}}

In this section, we derive the mean first passage time Eqs.~(\ref{tau_min}) and (\ref{tau_E0}).  Rather than solve Eq.~(\ref{ODE_tau}) in the text, it is simpler to find the transition time from starting point $x$ on energy branch $j=1,2$ to the saddle point $E_b$, then account for the probability to transition to the other branch, rather than return to the same well.  We therefore find the solution, $\tau_j(x)$, of
\beq\label{tau_ODE}\begin{aligned}
\big[-\alpha A_j(x) & + \aJ B_j(x) + 2\alpha\epsilon C_j(x)\big]\tau_j'(x) \\
& + \alpha\epsilon A_j(x)\tau_j''(x) = -1
\end{aligned}\eeq
with absorbing boundary condition $\tau_j(E_b)=0$, and divide it by the switching probability in Eq.~\eqref{Pswitch}.  First we consider the solution valid for any temperature, then consider the limit of vanishing temperature.

The exact solution to \eqref{tau_ODE} requires a second boundary condition.  As $x\to E_a$ we know from Sec.~\ref{sec:asym_min} that $A_j(x)\to 0$ and $B_j(x)\to 0$, which leaves the condition 
$$
2\alpha\epsilon C(E_a) \tau'(E_a)= -1.
$$ 
Using integrating factors, we integrate \eqref{tau_ODE} twice and obtain
\beq\label{tau_exact}\begin{aligned}
 \tau_j(x) =& \frac{\gamma_1+\gamma_2}{\gamma_j} \int_x^{E_b} \left( \frac{1}{2 \alpha\epsilon (\beta_y+\beta_z \pm h_x)} + \frac{1}{\alpha\epsilon} I_j(y) \right)\\
 & \times e^{(y-E_a+\frac{\aJ}{\alpha}F_j(y))/\epsilon-G_j(y)} dy
 \end{aligned}\eeq
where
$$ 
I_j(y) = \int_{E_a}^y \frac{1}{A(z)} e^{-(z-E_a+\frac{\aJ}{\alpha}F_j(z))/\epsilon+G_j(z)} dz 
$$
and where 
$$
F_j(z)=\int_{E_a}^z \frac{B_j(t)}{A_j(t)} dt\qquad\textrm{and}\qquad G_j(z)=\int_{E_a}^z \frac{2C_j(t)}{A_j(t)} dt.
$$
The expression for $\gamma_j$ was described in Appendix~\ref{sec:Matching}; it is
$$\begin{aligned}
\gamma_1 \approx 4 d_2 (\beta_z^2-\beta_y^2)-b_2(4\beta_y^2+h_x^2) \\
\gamma_2 \approx 4 d_1 (\beta_z^2-\beta_y^2)-b_1(4\beta_y^2+h_x^2)
\end{aligned}$$
where $b_j$ and $d_j$ are defined in Eq.~\eqref{bjS}.

For vanishing temperature ($\epsilon\to 0$), rather than approximate \eqref{tau_exact} directly, we notice that the solution has a boundary layer where the coefficients $A_j(x)$ and $B_j(x)$ go to zero: both near $x=E_{a,j}$, the minimum ($E_{a,1}=-h_x$ and $E_{a,2}=h_x$), and $x=E_b$, the saddle point in energy.  We match the solution coming out of the boundary layer near $E_b$ to determine the leading order expression for the mean first passage time  from the meta-stable point.  This meta-stable point is either $E_0>E_{a,j}$ for values of $\aJ$ when a stable limit cycle exits on branch $j$ or the minimum value, $E_{a,j}$.  For simplicity of notation, we will drop the subscript $j$ and only consider $j=1$.  The solution for $j=2$ is derived similarly.

First, we consider the boundary layer near $E_b$, and rescale the energy by $\epsilon$ defining $\xi=(x-E_b)/\epsilon$ so that $\xi\to -\infty$ leaves the boundary layer.  The rescaled equation for $g(\xi)=\tau(x(\xi))$ is
$$\begin{aligned}
\big[ -\alpha A(x(\xi)) & + \aJ B(x(\xi)) + 2\alpha\epsilon C(x(\xi)) \big] \frac{1}{\epsilon} g'(\xi) \\
& + \frac{\alpha\epsilon}{\epsilon^2} A(x(\xi)) g''(\xi) = -1
\end{aligned}$$
which to leading order reduces to
$$
\left[ -1 + \frac{\aJ}{\alpha} \frac{B(x(\xi))}{A(x(\xi))}  \right] g'(\xi) + g''(\xi) = 0
$$
with boundary condition $g(0)=0$.  We then have that
$$
g'(\xi) = c \exp\left[ \xi - \frac{\aJ}{\alpha} \int_0^\xi \frac{B(E_b+\epsilon z)}{A(E_b+\epsilon z)} dz \right]
$$
and integrating again yields
$$
g(\xi) = c \int_0^\xi \exp\left[ y - \frac{\aJ}{\alpha} \int_0^y \frac{B(E_b+\epsilon z)}{A(E_b+\epsilon z)} dz \right] dy
$$
where we have used the boundary condition $g(0)=0$.  By first expanding the integral in the exponent in term of $\epsilon$, 
$$
\int_0^y \frac{B(E_b+\epsilon z)}{A(E_b+\epsilon z)} dz = \int_0^y \frac{\alpha}{\lambda} + O(\epsilon) dz \sim \frac{\alpha}{\lambda}y,
$$
where $\lambda$ is defined in \eqref{eq:ac}, we have that
$$
g(\xi) = c \int_0^{\xi} e^{(1-\aJ /\lambda)y}dy = \frac{c}{1-\frac{\aJ}{\lambda}}\left( e^{(1-\aJ/\lambda)\xi} - 1\right)
$$
and therefore
\beq\label{tau_c}
\tau(x) \approx \frac{c}{1-\frac{\aJ}{\lambda}}e^{(1-\aJ/\lambda )(E_b-x)/\epsilon} 
\eeq
to leading order.  We are left to determine the constant $c$.  As we leave the boundary layer,
$$
g(-\infty) = c \int_0^{-\infty} e^{(1-\aJ /\lambda)y}dy = \frac{-c}{1-\frac{\aJ}{\lambda}},
$$
and we see the solution becomes constant.   We turn to consider the full solution in the outer region away from the boundary layer to match this constant.

Returning to Eq.~\eqref{tau_ODE}, and dividing by $\alpha\epsilon A(x)$, we have
$$
 \frac{-1}{\epsilon}\left( 1 - \frac{\aJ}{\alpha} \frac{B(x)}{A(x)} - \epsilon \frac{2C(x)}{A(x)} \right)\tau'(x) + \tau''(x) = \frac{-1}{\alpha\epsilon A(x)}.
$$
We rewrite this as
\beq\label{tau_1}
\left[ e^{\Phi(x)/\epsilon}\tau'(x)\right]' = \frac{-1}{\alpha\epsilon A(x)} e^{\Phi(x)/\epsilon}
\eeq
where 
$$\begin{aligned}
\Phi(x) &  \equiv \Phi_0(x) + \epsilon \Phi_1(x) \\
& = -x + \frac{\aJ}{\alpha} \int_{*}^x \frac{B(y)}{A(y)} dy + \epsilon \int_*^{E_b} \frac{2C(y)}{A(y)}dy 
\end{aligned}$$
for some arbitrary point $*$.
After integrating \eqref{tau_1} from $x_0$ to $E_b$ we have
$$
e^{\Phi(E_b)/\epsilon} \tau'(E_b) - e^{\Phi(x_0)/\epsilon}\tau'(x_0) = \frac{-1}{\alpha \epsilon} \int_{x_0}^{E_b} \frac{1}{A(y)} e^{\Phi(y)/\epsilon} dy.
$$
The constant from \eqref{tau_c}, enters through $\tau'(E_b) = g'(0)/\epsilon = c / \epsilon$.  Combining with the above equation we have
\beq\label{eq_c}\begin{aligned}
c =& \epsilon \tau'(x_0)e^{-(\Phi(E_b)-\Phi(x_0))/\epsilon} \\
& -\frac{1}{\alpha}\int_{x_0}^{E_b} \frac{1}{A(y)}e^{-(\Phi(E_b) - \Phi(y))/\epsilon} dy.
\end{aligned}\eeq

The integral in \eqref{eq_c} is dominated by what happens near $x_0$, and we have two cases, the first when $x_0$ is the solution to $-\alpha A(x)+\aJ B(x) = 0$ in which case $A(x_0)\ne 0$, and the point $x_0$ is away from either boundary layer.  The second is when $x_0$ is the minimum, and $1/A(x)$ must be canceled by the term generated from the integral of $C(x)/A(x)$ in $\Phi(x)$.  In either case, we will need the expansion of 
$$-(\Phi_0(E_b) - \Phi_0(x)) = E_b-x - \frac{\aJ}{\alpha} \int_x^{E_b} \frac{B(y)}{A(y)} dy $$
 in terms of $\epsilon$ defined by $\xi=(x-E_b)/\epsilon$.  We then have
$$\begin{aligned}
-(\Phi_0(E_b) - \Phi_0(x)) & = -\epsilon\xi - \frac{\aJ}{\alpha}[0 + \epsilon\xi (-1)\frac{\alpha}{\lambda} +O(\epsilon^2)     ] \\
& \sim -(1-\frac{\aJ}{\lambda} )\epsilon\xi.
\end{aligned}$$

When $x_0$ is the solution to $-\alpha A(x_0)+\aJ B(x_0) = 0$, the mean passage time, $\tau$, is approximately constant at $x_0$ and therefore $\tau'(x_0)\approx 0$.  The expansion of $-(\Phi_1(E_b) - \Phi_1(x)) $ only contributes higher order terms to the exponent, and
$$\begin{aligned}
&\frac{-1}{\alpha}\int_{x_0}^{E_b} \frac{1}{A(y)}e^{-(\Phi(E_b) - \Phi(y))/\epsilon} dy \\
&\approx  \frac{-1}{\alpha}\int_{-\infty}^{0} \frac{1}{A(x_0)}e^{-(1-\frac{\aJ}{\lambda} )\xi }\epsilon d\xi
= \frac{\epsilon}{\alpha}\frac{1}{A(x_0)(1- \frac{\aJ}{\lambda} )}.
\end{aligned}$$
This, together with $\tau'(x_0)=0$, gives the constant in Eq.~\eqref{tau_c}.  Combining with the switching probability factor, we have Eq.~(\ref{tau_E0}) in the text.

On the other hand when $x_0=E_a$, the expansion of $-(\Phi_1(E_b) - \Phi_1(x)) $ includes a large term near $E_a$.  From the scalings worked out in Sec.~\ref{sec:asym_min}, we know $2C(x)/A(x) \sim 1/(x-E_a)$, which produces a large term, $\log (x-E_a)$, in the expansion of $-(\Phi_1(E_b) - \Phi_1(x)) $.  Together with the scaling $A(x)\sim 2(\beta_y+\beta_z+h_x)(x-E_a)$ near $E_a$, we have
$$\begin{aligned}
-\frac{1}{\alpha}&\int_{E_a}^{E_b}  \frac{1}{A(y)}e^{-(\Phi(E_b) - \Phi(y))/\epsilon} dy  \\
& \approx -\frac{1}{\alpha}\int_{E_a}^{E_b} \frac{e^{-(\Phi_0(E_b) - \Phi_0(y))/\epsilon+\log (x-E_a)}}{2(\beta_y+\beta_z+h_x)(x-E_a)} dy \\
& \approx -\frac{1}{\alpha}\int_{-\infty}^{0} \frac{e^{-(1-\frac{\aJ}{\lambda} )\xi }}{2(\beta_y+\beta_z+h_x)} \epsilon d\xi
\\&=  \frac{\epsilon}{\alpha}\frac{1}{2(\beta_y+\beta_z+h_x)(1- \frac{\aJ}{\lambda} )}.
\end{aligned}$$
For the term in \eqref{eq_c} involving $\tau'(E_a)$, we return to Eq.~\eqref{tau_ODE}, where for $x\ll \epsilon$ we have
$$ 2\alpha\epsilon C(x)\tau'(x) = -1 $$
to leading order and therefore
$$ \tau'(E_a) = \frac{-1}{ 2\alpha\epsilon C(E_a)} .$$
We then have
$$ \begin{aligned}
\tau'&(E_a)e^{-(\Phi(E_b)-\Phi(E_a))/\epsilon} \\
&\approx \lim_{x\to E_a} \frac{1}{2\alpha\epsilon C(x)} e^{ -(\Phi_0(E_b)-\Phi_0(x))/\epsilon + \log (x-E_a)} 
\end{aligned}$$
to leading order in the exponent, which goes to zero due to the $\log(x-E_a)$ term.  Thus, the $\tau'(E_a)$ term does not contribute to the solution.  Combining with the switching probability factor, we have Eq.~(\ref{tau_min}) in the text.

 \bigskip



%

\end{document}